\documentclass[iop]{emulateapj}

\usepackage{amsmath}
\usepackage{amsfonts}
\usepackage{bm}
\usepackage{color}
\usepackage{graphicx}
\definecolor{darkgreen}{cmyk}{0.85,0.2,1.00,0.2}

\usepackage{amsbsy}
\usepackage{hyperref}

\usepackage{amsmath,amssymb,subfigure}
\usepackage{graphicx}
\usepackage{epsfig}
\usepackage{color}
\usepackage{ulem}
\usepackage{epstopdf}
\newcommand{\be}{\begin{eqnarray}}

\newcommand{\ee}{\end{eqnarray}}

\newcommand{\bxe}{\bar{x}_e}
\newcommand{\bxh}{\bar{x}_H}
\newcommand{\cdd}{\xi_{\delta\delta}}
\newcommand{\cdx}{\xi_{x\delta}}
\newcommand{\cxx}{\xi_{xx}}
\newcommand{\ls}{\mathrel{\raise0.27ex\hbox{$<$}\kern-0.70em \lower0.71ex\hbox{{$\scriptstyle \sim$}}}}

\begin{document}

\title{Probing patchy reionization through $\tau$-21cm correlation statistics}

\author{P.~Daniel Meerburg$^{1}$}\email{meerburg@princeton.edu}
\author{Cora Dvorkin$^{2}$}\email{cdvorkin@ias.edu}
\author{David N. Spergel$^{1}$}\email{dns@astro.princeton.edu}
\affiliation{$^1$Department of Astrophysical Sciences, Princeton University, Princeton, NJ 08540 USA. }
\affiliation{$^2$ Institute for Advanced Study, Einstein Drive, Princeton, NJ 08540, USA}
\date{\today}

\begin{abstract} 
We consider the cross-correlation between free electrons and neutral hydrogen during the epoch of reionization. The free electrons are traced by the optical depth to reionization $\tau$ while the neutral hydrogen can be observed through $21$ cm photon emission. As expected, this correlation is sensitive to the detailed physics of reionization. Foremost, if reionization occurs through the merger of relatively large halos hosting an ionizing source, the free electrons and neutral hydrogen are anti-correlated for most of the reionization history. A positive contribution to the correlation can occur when the halos that can form an ionizing source are small.  A measurement of this sign change in the cross-correlation could help disentangle the bias and the ionization history. We estimate the signal-to-noise of the cross-correlation using the estimator for inhomogeneous reionization $\hat{\tau}_{\ell m}$ proposed by Dvorkin and Smith (2009). We find that with upcoming radio interferometers and CMB experiments, the cross-correlation is measurable going up to multipoles $\ell\sim 1000$. We also derive parameter constraints and conclude that, despite the foregrounds, the cross-correlation proofs a complementary measurement of the EoR parameters to the 21 cm and CMB polarization auto-correlations expected to be observed in the coming decade.
\end{abstract}

\maketitle
\section{Introduction}
The epoch of reionization (EoR) is one of the least understood periods of cosmic history, with only limited observational measurements (see e.g. \cite{Loeb:2000fc}). 
The  absence of the Gunn-Peterson trough in the spectra of quasars implies that reionization should have been completed around $z=6$ \citep{Fan:2005es,Fan:2006dp}. The total optical depth to reionization has been measured to be $\tau = 0.084 \pm 0.013$ \citep{Hinshaw:2012fq}. If reionization is assumed to be instantaneous, this would imply a transition redshift of $z_{\rm re}=11$. Besides these constraints, we know very little about the details of reionization, such as the typical halo mass associated with the first ionizing objects as well as the distribution of these objects inside the halos.  

The spectral mapping of neutral hydrogen in emission \citep{1979MNRASHR,1990MNRAS.247..510S,McQuinn:2005hk, 1997ApJ...475..429M,Zaldarriaga:2003du,Furlanetto:2004ha,Furlanetto:2004nh,Furlanetto:2006jb} promises to be a new probe of the EoR. The spontaneous hyperfine spin flip transition causes the emission of a photon with a wavelength of 21 centimeters in the rest frame. Applying different base filters to the observed emission, it is possible to map the distribution of neutral hydrogen in the Universe as a function of redshift. The auto-correlation of the observed maps is very sensitive to the EoR parameters. 

Additionally, cross correlating the observed maps with other observables could provide complementary constraints on the EoR parameters. For example, the 21 cm fluctuations are expected to be correlated with galaxies  \citep{Lidz:2008ry,Wiersma:2012ek} as well as with the Cosmic Microwave Background (CMB) fluctuations through the Doppler peak and the kinetic SZ effect (see e.g. \cite{Alvarez:2005sa, 2005MNRAS.360.1063S,Adshead:2007ij,2008MNRAS.389..469T,Tashiro:2009uj,Tashiro:2010fj,Holder:2006qi} and \citep{2010MNRAS.402.2279J,Natarajan:2012ie} for recent simulations). 
Unlike the fluctuations in 21 cm, which are a direct representation of the underlying neutral hydrogen, no direct measurement of the electron density at high redshifts exists. The electron density can be measured indirectly through its integrated effect on the CMB, providing us with a number ($\tau$) that tells us the fraction of photons affected by scattering of electrons along the line of sight. One can go further and reconstruct the inhomogeneities in the optical depth field by considering second order effects on the CMB due to the screening mechanism \citep{Dvorkin:2009ah}, Thomson scattering and the kSZ effect (see \cite{Dvorkin:2008tf}, where an estimator of the anisotropic optical depth field is derived and \cite{Gluscevic:2012qv} for an implementation of this estimator to WMAP $7$-year data). In this work we consider the cross-correlation between a reconstructed map of the inhomogeneous optical depth $\tau_{\ell m}$ (using CMB polarization observations) and a neutral hydrogen map measured through the redshifted 21 cm lines. Intuitively, these two observables are expected to be anti-correlated on most scales. We would like to stress that this cross-correlation is fundamentally different from direct cross correlations between CMB temperature and polarization and 21 cm maps \citep{2008MNRAS.389..469T,Tashiro:2009uj}; the $\tau$ estimator is a quadratic estimator, making this cross correlation a statistical 3 point correlation function rather than a 2 point function. 

Besides providing complementary constraints on the EoR parameters, the cross-correlation between the 21 cm field and the CMB should in principle be less sensitive to the details of the foregrounds. Although current \citep{Masui:2012zc,Chapman:2012pn} and upcoming experiments \citep{Harker:2010ht, Mellema:2012ht} are expected to be capable of measuring the auto-correlation of 21 cm maps, one very persistent nuisance in extracting the signal from reionization are the foregrounds. Inhomogeneities in the 21 cm signal due to patchy reionization must be separated from fluctuations in foreground sources.  Typical foreground sources are faint radio galaxies, starburst galaxies and galaxies responsible for 
reionization. In addition, our own Galaxy is very bright at the frequencies one aims at for mapping the 21 cm signal from reionization, exceeding the 21 cm reionization signal by several orders of magnitude. Attempts have been made to characterize these foregrounds \citep{Liu:2009qga,2008MNRAS.389.1319J,Bernardi:2010xn,Bernardi:2009pi,Liu:2012xy}. Despite these efforts, foregrounds can never be fully removed, simply because we do not know their exact origin. 

This paper is organized as follows. We review the physics of reionization and derive the expressions for the fluctuations in 21 cm $a_{\ell m}^{21}$ and fluctuations in the optical depth $\tau_{\ell m}$ in \S\ref{sec:21cm_reionization}. The former is proportional to the neutral hydrogen fraction, while the latter is proportional to the free electron fraction. Using a simple reionization model \citep{Furlanetto:2004nh,Wang:2005my} where regions of HII are represented by spherical bubbles of typical size $\bar{R}$ we derive an expression for the cross-correlation $\langle \tau_{\ell m} a_{\ell m}^{21 *}\rangle$ in section  \S\ref{sec:cross_correlation}. In \S\ref{sec:2bubble} and \S\ref{sec:1bubble} we study the one-bubble and two-bubble contributions to the power spectrum. We compute the angular power spectrum of the cross-correlation in \S\ref{sec:correlation_2D}. We assess the observability of the cross correlation by using a redshift weighting to maximize the signal to noise. 
We end this section with an estimate of the EoR parameter constraints, when considering LOFAR and SKA noise levels.  We present our conclusions in \S\ref{sec:Discussion and Conclusion}.  In the Appendix we discuss the dependence of the cross-correlation on the parameters of the reionization model.  

Unless specifically mentioned, we use the following set of parameter values throughout the paper: $h=0.704, \Omega_b=0.044, \Omega_c=0.23, \Omega_{K}=0, n_s=0.96$ and $\tau=0.084$. Lensing is included and we use the non-linear halo fit model to determine the power spectrum of density fluctuations. We use the WMAP pivot scale $k_* = 0.002$ Mpc$^{-1}$ and $A_s=2.46\times10^{-9}$ \citep{Hinshaw:2012fq}. 

\section{21 cm brightness and the optical depth to reionization}
\label{sec:21cm_reionization}

In this section we will review the standard results for fluctuations in the 21 cm temperature brightness and relate those to fluctuations in the neutral hydrogen fraction. In the second half of this section, we will derive the fluctuations in the optical depth $\tau$ caused by fluctuations in the free electron fraction along the line of sight, confirming the results first obtained in \citep{Holder:2006qi}. 
 
 The optical depth of a region of the IGM in the hyperfine transition is given by \citep{Field:1958}
 \be
 \tau_{21}(z) = \frac{3c^3 h A_{10} }{32 \pi k \nu_0^2 T_S} \frac{n_{HI}}{(1+z) (dv_{\parallel}/dr_{\parallel})},
 \ee
where $\nu_0 = 1420.4$ MHz is the rest frame hyperfine 21 cm ($\nu_0 = \lambda_{21}/c$) transition frequency,  $A_{10} = 2.85\times10^{-15} \mathrm{s}^{-1}$ is the spontaneous emission coefficient for this transition, $T_S$ is the spin temperature of the IGM, weighting the relative population of the atoms in the singlet state to atoms in the triplet state \citep{Field:1958}, $n_{HI}$ is the neutral hydrogen density, and $v_{\parallel}$ the proper velocity along the line of sight. At high redshifts, where peculiar motions along the line of sight are small compared to the Hubble flow, $dv_{\parallel}/dr_{\parallel}=H(z)/(1+z)$. At $z=10$ dark energy and radiation are both unimportant and we can solve for $H(z)$ in a matter dominated Universe, $H(z)\simeq H_0 \Omega_m^{1/2} (1+z)^{3/2}$.

We can now write the following expression for the optical depth:
 \be\label{eq:tau21}
 \tau_{21}(z)&\simeq& 8.6\times 10^{-3} (1+\delta_b) x_H\left[\frac{T_{cmb}}{T_S}\right]\left[\frac{1-Y_p}{1-0.248}\right]\times\nonumber\\
 && \left(\frac{\Omega_b }{0.044}\right)\left[\left(\frac{0.27}{\Omega_m}\right)\left(\frac{1+z}{10}\right)\right]^{1/2}
 \ee
 Here we used $T_{cmb} = 2.73 (1+z) \mathrm{K}$, $\delta_b=(\rho_b-\bar{\rho_b})/\bar{\rho_b}$ and
 \be
 n_{HI} &\simeq& (1-Y_p)x_H \frac{\Omega_b}{\Omega_m} \frac{\rho_m}{m_p},\nonumber
 \ee
where $x_H$ is the neutral hydrogen fraction, i.e. $x_H = n_{HI}/(n_{HI}+n_e)$, $\rho_m$ is the matter energy density and $m_p$ is the proton mass. The factor $(1-Y_p)$ addresses the fact that not all protons are in hydrogen, but a fraction is in Helium. 

The intensity along the line of sight from a thermal source is given by 
\be
I&=& I_0 e^{-\tau} + \int_{\tau}^0 d\tau' e^{-\tau'} \frac{\eta_{\nu}}{\kappa_{\nu}},
\ee
with $\kappa_{\nu}$ the absorption coefficient and $\eta_{\nu}$ the emissivity of photons. Using $dI  = \eta_{\nu} dl -\kappa_{\nu} I dl = 0$ in the Rayleigh-Jeans limit, we have $I = 2kT_b \nu^2/c^2$, while $\eta_{\nu}/\kappa_{\nu} = 2kT_S\nu^2/c^2$ and $I_0 = 2kT_{cmb}\nu^2/c^2$. Hence, we can write the 21 cm brightness temperature as:
\be
T_b = T_{cmb}e^{-\tau_{21}}+T_S(1-e^{-\tau_{21}})
\ee

The brightness temperature increment is defined at an observed frequency $\nu$ corresponding to a redshift $1+z= \nu_0/\nu$ as
\be\label{eq:deltaTb}
\delta T_b(z) \equiv \frac{T_b-T_{cmb}}{1+z} \simeq \frac{(T_S-T_{cmb})}{1+z} \tau_{21}
\ee

Using Eq. \eqref{eq:tau21}, we can re-write Eq. \eqref{eq:deltaTb} as \citep{1990MNRAS.247..510S,1997ApJ...475..429M}
\be
\delta T_b(z)&\simeq& 27\; \mathrm{mK}\; (1+\delta_b)x_H \left[\frac{T_S-T_{cmb}}{T_S}\right]\left[\frac{1-Y_p}{1-0.248}\right]\nonumber\\
 &&\times\left(\frac{\Omega_b}{0.044}\right)\left[\left(\frac{0.27}{\Omega_m}\right)\left(\frac{1+z}{10}\right)\right]^{1/2}
\ee

There are usually two types of filters associated with the resolution of the experiment. First, there is a finite angular resolution, which will affect all modes perpendicular to the line of sight. Second, since the brightness temperature of the 21 cm emission is a 3-dimensional field, we are confined to a frequency resolution or bandwidth, which affects the modes along the line of sight (or, equivalently, a redshift resolution). 

The total integrated 21 cm surface brightness is given by
\be
T_b(\hat{n},\chi)&=&T_0(\chi)\int d\chi' W_{\chi}(\chi') \psi(\hat{n},\chi ')
\ee

Here $W_{\chi}$ is an experimental band filter that is due to the finite frequency resolution of the instrument, which is centered around $\chi$ (comoving distance).
We define the dimensionless brightness temperature $\psi$ as
\be
\psi = (1+\delta_b)x_H\left({T_S-T_{cmb}\over T_{S}}\right)
\ee
In the limit of $T_s\gg T_{cmb}$, $\psi=(1+\delta_b)x_H$. 

Now $T_0(z)$ can be written as 
\be
T_0(z) & \simeq & 27\; \mathrm{mK}\left[\frac{1-Y_p}{1-0.248}\right] \nonumber \\
&&\left(\frac{\Omega_b}{0.044}\right)\left[\left(\frac{0.27}{\Omega_m}\right)\left(\frac{1+z}{10}\right)\right]^{1/2}
\ee

We will now consider fluctuations in the free electron density, which in turn will induce fluctuations in the optical depth. The optical depth to distance $\chi$ along the line of sight is given by
\be
\tau(\hat{n},\chi)&=&\sigma_T \int_0^\chi d\chi' n_e(\hat{n},\chi ') a(\chi'),
\ee
where $\sigma_T$ is the Thomson cross-section, $n_e$ is the electron number density and $a$ is the scale factor. Relating the free electron density to the free electron fraction $x_e$, we can write 
\be
n_e(\hat{n},\chi)&\simeq& \frac{x_e \rho_b}{m_p}(1-{3\over 4}Y_p),
\ee
assuming that Helium is singly ionized.

The average baryon density diffuses in an expanding background as $a^{-3}$, and the free electron density becomes
\be
n_e(\hat{n},\chi)&=& (1-{3\over 4}Y_p)\frac{ \rho_{b,0}}{m_p} a^{-3} (1+\delta_b)x_e
\ee

The optical depth can in turn be written as
\be
\tau(\hat{n},\chi)&=& \sigma_T (1-{3\over 4}Y_p)\frac{\rho_{b,0}}{m_p}  \int_0^{\chi}\frac{d\chi'}{a^2(\chi')} x_e(\hat{n},\chi')\nonumber\\
&\times&(1+\delta_b(\hat{n},\chi'))
\ee 

Therefore, we can relate fluctuations in the optical depth $ \delta\tau$ to fluctuations in the 21 cm brightness temperature $\delta T_b$ \citep{Holder:2006qi} as
\be\label{eq:deltatau_21}
\delta \tau &=&(1-{3\over 4}Y_p) \frac{\sigma_T \rho_{b,0}}{m_p H_0 \Omega_m^{-1/2}} \int dz \left[(1+z)^{1/2} \delta_b -\frac{\delta T_b(z)}{8.5 \mathrm{mK}} \right],\nonumber\\
\ee
where we have assumed a delta window function. 

It is worth noticing that the above expression for the fluctuations in the 21 cm brightness  temperature is only valid for  $T_s>T_{\rm CMB}$. Early on, when the number of ionizing sources are rare and the temperature of the IGM close to these sources is coupled to the kinetic temperature by Ly$\alpha$ photons associated with these local sources, this assumption breaks down, and the 21 cm signal can appear in absorption. We neglect this effect in this paper. 

\section{Correlating $X$ and $\psi$}
\label{sec:cross_correlation}

We will now cross-correlate the optical depth fluctuations with the temperature brightness.
As we saw before, the CMB optical depth is proportional to the free electron density. If reionization is inhomogeneous,  the free electron density is a function of position in the sky.
Anisotropies in the optical depth produce three effects in the CMB:  (i) screening of the temperature and polarization fluctuations that we observe today by an overall factor of $e^{-\tau(\hat{n})}$. This effect generates CMB B-mode polarization; (ii) Thomson scattering: new polarization is generated by scattering of the local temperature quadrupole that each electron sees along the line of sight. This effect also produces B-modes; and (iii) new temperature anisotropy is generated from the radial motion of ionized bubbles relative to the observer (the kinetic Sunyaev Zel'dovich effect).

The two-point correlation function between the E-modes and the B-modes generated from patchy reionization is proportional to the anisotropic part of the optical depth. This fact allowed the authors in Ref. \citep{Dvorkin:2008tf} to write a minimum variance quadratic estimator $\hat{\tau}_{\ell m}$ for the field $\tau(\hat{n})$. In this work, we will use the CMB polarization fluctuations to reconstruct a map of $\tau$ and cross-correlate it with the 21 cm field.

We will use the shorthand notation  $X(\hat{n},\chi) = x_e(1+\delta_b)$. 
We can write the cross-correlation between the field X (measured through the CMB) and the field $\psi$ (measured through 21 cm) as
\be\label{eq:cross_corr}
\xi_{X\psi}&\simeq& -\xi_{xx}(1+\xi_{\delta\delta})-(\bxh-\bxe+\xi_{x\delta})\xi_{x\delta} \nonumber \\
&&  + (\bxh-\bxh^2)\xi_{\delta\delta} 
\ee

Here we defined $\cxx = \langle x_H(\vec{x}_1)x_H(\vec{x}_2)  \rangle -\bxh^2$, $\cdd = \langle \delta_b (\vec{x}_1)\delta_b(\vec{x}_2)  \rangle $ and $\cdx = \langle \delta_b (\vec{x}_1) x_H(\vec{x}_2)  \rangle$. We make the simplistic assumption that the connected part of $\langle \delta_x\delta_x\delta_b\delta_b\rangle$ vanishes. Here $\delta_x$ corresponds to fluctuations in the neutral hydrogen fraction, which is given by $x_H=\bar{x}_H\left(1+ \delta_x\right)$. 

Before we can compute $\xi_{X\psi}$ we need to specify our model of reionization. We will assume that the Universe reionized through the growth of ionized bubbles associated with massive halos. The bubbles themselves contain a single source and we assume their size to be larger than the non-linear scale.  

We will adopt the following average reionization fraction as a function of redshift
\be\label{eq:xe}
\bxe(z)&=& \frac{1}{2} \left[1+\tanh \left(\frac{y_{\rm re}-(1+z)^{3/2}}{\Delta y}\right)\right],
\ee
which is the one used in the code CAMB \citep{Lewis:1999bs}. Here $y(z)=(1+z)^{3/2}$, $y_{\rm re}=y(z_{\rm re})$ and $\Delta y$ are free parameters that satisfy our integrated optical depth along the line of sight $\tau=0.084$. 

The ionized bubble around a given source is assumed to be spherical with an average radius $\bar{R}$.
We will assume that the typical ionized bubble radii are log-normal distributed \citep{Zahn:2006sg}, i.e. there is a skewness towards smaller bubble sizes, 
\be
P(R) = \frac{1}{R} \frac{1}{\sqrt{2\pi\sigma^2_{\ln R}}} e^{-[\ln (R/\bar{R})]^2/(2\sigma_{\ln R}^2)},
\ee
where  $\sigma_{\ln R}$ is the variance of the distribution. 

The average bubble volume is then given by 
\be
\langle V_b \rangle &=&\int dR P(R) V_b(R) = \frac{4\pi \bar{R}^3 }{3} e^{9\sigma_{\ln R}^2/2}
\ee
Hence, we can  define a volume weighted radius, $R_0$ such that $\langle V_b \rangle = 4\pi R_0^3/3$, which can be written as
\be
R_0 = \bar{R}e^{3\sigma_{\ln R}^2/2}
\ee
 
If we assume that a given point in space is ionized with Poisson probability, we can write the ionization fraction as
\be
\langle x_e(\vec{x}) \rangle_P = 1- e^{-n_b(\vec{x}) \langle V_b \rangle}, 
\label{eq:xe}
\ee
with $n_b$ the number density of bubbles. The brackets around $x_e$ are placed to remind us that we are considering a Poisson distribution of sources, and the result is averaged over the Poisson process. We further assume that the number density of bubbles traces the large-scale structure with some bias $b$:
\be
n_b(\vec{x}) &=& \bar{n}_b\left(1+b \delta_W(\vec{x})\right),
\label{eq:nb}
\ee
while the average bubble number density is related to the mean ionization fraction as
\be
\bar{n}_b &=& -\frac{1}{\langle V_b \rangle} \ln (1-\bxe)
\label{eq:barnb}
\ee
Here  $\delta_W$ is the matter over-density $\delta$ smoothed by a top hat window of radius $R$,
\be
\delta_W(\vec{x}) &=& \int d^3 x' \delta(\vec{x}')W_R(\vec{x}-\vec{x}')
\ee
In momentum space,
\be
W_R(k) = \frac{3}{(kR)^3}\left[\sin (kR)-kR\cos(kR)\right],
\ee
which is  the Fourier transform of $W_R(x) = V_b^{-1}$ for $x \leq R$ and $W_R(x) = 0$ otherwise.

We further define 
\be
\langle W_R\rangle(k)&=& \frac{1}{\langle V_b\rangle} \int_0^{\infty} dR P(R) V_b(R) W_R(k R)
\label{eq:Wangle}
\ee
and 
\be
\langle W_R^2\rangle(k)&=& \frac{1}{\langle V_b\rangle^2} \int_0^{\infty} dR P(R) \left[V_b(R) W_R(k R)\right]^2
\label{eq:W2angle}
\ee

\section{Reionization parameters}\label{sec:reionization_model}

Reionization can only proceed if the seed halo is massive enough for cooling. In particular, line cooling and atomic cooling are important for collapse. One can relate the virial temperature of the halo to the mass of the halo 
\be
\frac{T_{\mathrm{vir}}}{10^4\mathrm{K}} &=& 1.1 \left(\frac{\Omega_m h^2}{0.15}\right)^{1/3}\left(\frac{1+z}{10}\right)\left(\frac{M}{10^8 M_\odot}\right)^{2/3}.
\label{eq:tvir}
\ee
Therefore, setting a condition on the amount of cooling necessary to form ionizing objects sets a typical mass of the halo, which will be redshift dependent. We can relate the mean number density of bubbles, the average reionization fraction and the typical bubble volume by inverting Eq. \eqref{eq:barnb}:
\be
\langle V_b \rangle &=& - \frac{1}{\bar{n}_b} \ln (1-\bar{x}_e)
\label{eq:Vb}
\ee
The mean bubble number density introduced in Eq.~\eqref{eq:barnb} is derived through an integral over the halo mass function, with a mass threshold $M_{\mathrm{th}}$ which can roughly be set by the viral temperature in Eq.~\eqref{eq:tvir} \footnote{Although the model we are using here is self consistent, relating our toy-reinoization model to all other relevant parameters, we find that the resulting bubble radius as a function of redshift is too small compared to simulations \citep{2008ApJ...681..756S}. One can alleviate this discrepancy somewhat by raising $M_{\mathrm{th}}$. For that purpose we assume the critical temperature to form an ionizing object to be five times the virial temperature.}:
\be
\bar{n}_b &=& \int^{\infty}_{M_{\mathrm{th}}} \frac{dn_h}{d\ln M} \frac{dM}{M}
\ee 
We use the Sheth and Tormen \citep{Sheth02a} halo mass function
\be
 \frac{dn_h}{d\ln M} &=& \frac{\rho_{m(0)}}{M} f(\nu) \frac{d\nu}{d \ln M}
\ee
with 
\be
\nu f(\nu) &=& A \sqrt{\frac{2}{\pi}a\nu^2} \left(1+(a\nu^2)^{-p}\right) e^{-a\nu^2/2}
\ee
where $\nu = \delta_c/\sigma_{lin}(M,z)$ and $\sigma_{lin}$ is the variance of the density field smoothed with the top-hat window function enclosing a mass $M$:
\be
\sigma_{lin}^2(M,z) &=& \int \frac{dk}{k} \Delta_m^2(k,z) W_{R(M)}^2(k)
\ee
It is straightforward to show that 
\be
 \frac{d\nu}{d \ln M} &=&  -\nu \frac{d \ln \sigma_{\mathrm{lin}}(M,z)}{d\ln M}
\ee
The parameters $\delta_c$, $a$ and $p$ can be fitted from simulations. $A$ is then derived through the constraint $\int d\nu f(\nu)=1$. 
Consequently, by setting $M_{\mathrm{th}}$ we can find an expression for the average bubble volume, and we can infer a function of the average bubble radius as a function of redshift (in the assumption of a log normal distribution of radii at any given redshift). 

Likewise, the bubble bias can be related to the halo bias (see e.g. \cite{Wang:2005my}) as:
\be
b = \frac{1}{\bar{n}_b} \int ^{\infty}_{M_{\mathrm{th}}} b_h(M) \frac{dn_h}{d\ln M} \frac{dM}{M}
\label{eq:bubblebias}
\ee
The integral runs over all masses with a threshold mass scale $M_{\mathrm{th}}$. Sheth and Tormen can be used for the halo bias
\be
b_h = 1+\frac{a\nu^2-1}{\delta_c}+\frac{2p}{\delta_c(1+(a\nu^2)^p)}
\ee
Therefore, for any given model of $\bar{x}_e(z)$ we can compute $\bar{R}(z)$, $b(z)$ and $\bar{n}_b(z)$. In this paper we will use $\delta_c = 1.686$, $a=0.707$ and $p=0.3$ \citep{2007MNRAS.374....2R}. 

For simplicity we will assume that $\sigma_{\ln R}$ is constant. Assuming a log normal distribution, for any given combination of $kR(z)$ we can then read of the value of $\langle W_R\rangle$ and $\langle W_R^2\rangle$ (see Figs.~\ref{fig:windowav} and ~\ref{fig:windowvar})

\begin{figure}
\centering
\includegraphics[width=.47\textwidth]{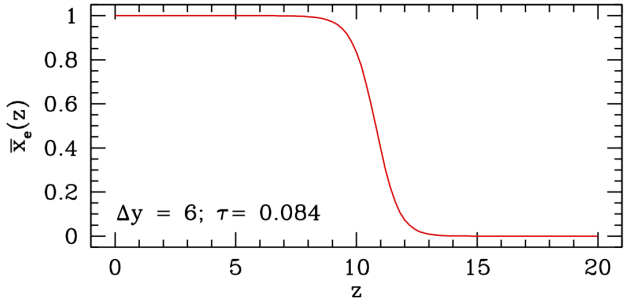}
\caption{The  average ionization fraction $\bxe(z)$ as a function of redshift $z$.}
\label{fig:xebar}
\end{figure}

In Fig.~\ref{fig:xebar} we have plotted the model we use for our average ionization fraction as a function of redshift. Our choice of parameters corresponds to a scenario with a neutral Universe at $z\geq 13$ and a completely ionized Universe at $z\leq 6$. 

\section{Two-bubble correlations}
\label{sec:2bubble}

The cross-correlation between the neutral hydrogen and free electron fraction has two main contributions. First, the correlation is set by the Poisson distribution of ionizing sources inside the bubbles. This term is referred to as the one-bubble term, and it is dominated by the shot noise. Second, cross correlations can also be induced by the enhanced probability of bubble formation (or ionizing sources) inside overdense regions (with probability $\langle x_e \rangle$). This term is referred to as the two-bubble term, and is relevant for scales much larger than the average size of a bubble.  


Using Eqs.~\eqref{eq:xe} and ~\eqref{eq:nb} we can Taylor expand $\langle x_e \rangle$ around small overdensities, to find
\be
\langle x_e \rangle &=& 1- e^{\ln (1-\bxe)(1+b\delta_W)}\nonumber \\
&\simeq &1 - (1-\bxe)\left[1+ b \delta_W \ln (1-\bxe)\right]+\mathcal{O}(\delta_{W}^2)\nonumber\\
\ee
with $b$ the bubble bias introduced earlier. 

Taking the Fourier transform of $\xi_{xx}$ and $\xi_{x\delta}$, we obtain
\be
P^{2b}_{xx}(k)&=&\left[\bxh\ln(\bxh)b \langle W_R\rangle(k)\right]^2P_{\delta\delta}(k)\\
P^{2b}_{x\delta}(k)&=& \bxh\ln(\bxh)b \langle W_R \rangle(k)P_{\delta\delta}(k),\label{eq:Pxd}
\ee
Note that here we are considering correlations of the neutral hydrogen fraction (perturbing the free electron fraction accounts for a minus sign in Eq. \eqref{eq:Pxd}).
The superscript ``$2b$" denotes the two-bubble contribution. Also, we assume that the baryon fluctuations (the gas) trace the dark matter fluctuations. 
The total two-bubble contribution to the power spectrum of $X\psi$ results in
\be
P^{2b}_{X\psi}(k) &\approx& - \bxh^2 [\ln \bxh b \langle W_R \rangle(k) +1]^2 P_{\delta\delta}(k) +\nonumber \\
&& \bxh [\ln \bxh b \langle W_R \rangle(k) +1] P_{\delta\delta}(k) 
\ee
Let us write $P^{2b}_{X\psi}(k)=Q(1-Q)P_{\delta\delta}(k)\equiv b_{\mathrm{eff}}P_{\delta\delta}(k)$, with $Q(k,z)= \bxh \ln \bxh b \langle W_R \rangle +\bxh$. We can distinguish two limiting cases: for $Q > 1$ the power spectrum effective bias $b_{\mathrm{eff}}$ is negative, representing an {\it anti-correlation}, while for $Q <1$, $b_{\mathrm{eff}}$ is positive, and the two-bubble term is positively correlated.  

\begin{figure}
\centering
\includegraphics[width=.47\textwidth]{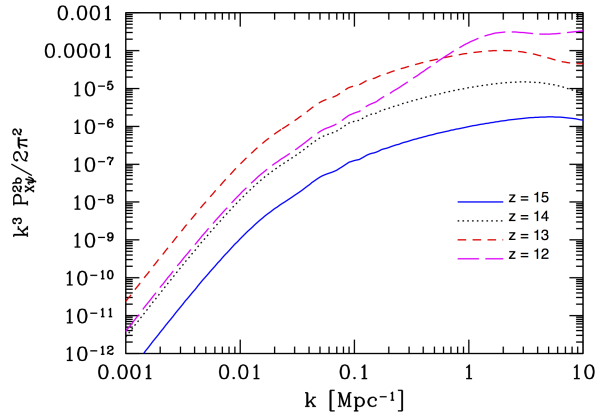}
\caption{$P^{2b}_{X\psi}(k) $ for 4 different redshifts with $\sigma_{\ln R}=0.5$. In our toy model the two-bubble term is positively correlated for all redshifts $z>12$. The zero point is uniquely determined by the relations $\bxh = e^{-1/b}$. Measuring this zero point could therefore help decorrelate these two parameters. However, the contribution of the two-bubble term to the overall cross correlation very small, and it will be be challenging to observe the cross correlation as a function of redshift as we show in section \S\ref{sec:correlation_2D}. }
\label{fig:p2_k}
\end{figure}

In the large scale limit, when $k\bar{R}\ll1$,  $b_{\mathrm{eff}} \rightarrow \bxh(b \ln \bxh + 1)(\bxe - b \ln\bxh)$.  
This function changes sign when $\bxh = e^{-1/b}$. The bubble bias on average grows towards larger $z$, despite the bubbles being smaller, the bubbles become rare (larger $\bxh$) and highly correlated. For our choice of parameters we can solve this equation for the redshift and find $z=12$ as the redshift at which the two-bubble term turns negative on large scales. Fig.~\ref{fig:p2_k} shows the two-bubble term for various redshifts.

\begin{figure}
\centering
\includegraphics[width=.47\textwidth]{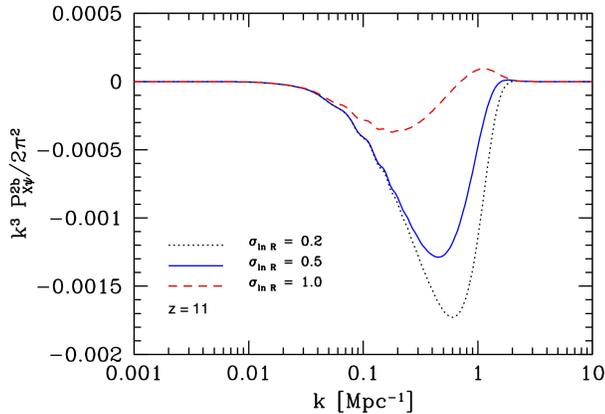}
\caption{ The two-bubble term for 3 different values of the width of the log-normal distribution. The relative contribution of the two-bubble term to the total correlation increases rapidly with decreasing $\bar{R}$ and $\sigma_{\ln R}$, as derived in the Appendix.  
}
\label{fig:P2bubble}
\end{figure}

At scales that are smaller than the radius of the bubbles, the two-bubble term is ill-defined \citep{2013arXiv1305.2917B}: the correlation length becomes shorter than the size of the bubbles, effectively rendering them to one bubble. Therefore, we will neglect the two-bubble correlation term at those scales. In practice, we apply a smoothing filter that effectively cuts the correlation for $kR_0(z) < 3$.
We show the two-bubble contribution to the cross-correlation in Fig. \ref{fig:P2bubble}.


\section{One-bubble correlations}
\label{sec:1bubble}

For scales much smaller than the average bubble radius, the correlation is dominated by the presence (or absence) of a single bubble \citep{Wang:2005my}. The correlation between two (ionized) points separated by $x_{12} = |\vec{x}_1-\vec{x}_2|$ can be written as \citep{Zaldarriaga:2003du, Furlanetto:2004ha}
\be
\langle x_e(\vec{x}_1) x_e(\vec{x}_2)\rangle &=& \bxe^2+(\bxe-\bxe^2)f(x_{12}/R),
\label{eq: 2p1bubble}
\ee
where $f(x)$ is a function with the following limits: $f(x)\rightarrow 1$ for $x\ll 1$ and $f(x)\rightarrow 0$ for $x\gg1$. If the probability for finding one point inside an ionized bubble is $\bxe$, then when $x_{12}\ll R$ the probability of finding the second point in the same bubble is $1$, hence their joint correlation probability is $\bxe$. For large separations, the probability of finding two points in separate bubbles is the product of both probabilities, i.e., $\bxe^2$.   Eq.~\eqref{eq: 2p1bubble} effectively encodes the smooth transition between these two regimes. The one-bubble correlation for free electrons (or equivalently neutral hydrogen) then becomes:
\be
\xi^{1b}_{x_ex_e} = \langle x_e(\vec{x}_1) x_e(\vec{x}_2)\rangle-  \bxe^2= (\bxe -\bxe^2)f(x_{12}/R)
\ee

As long as the bubbles do not overlap, the function $f$ can be described by the convolution of two top hat window functions $\langle W_{R}^2 \rangle$. The one-bubble contribution to the power spectrum can then be written as
\be
P^{1b}_{X\psi}&=&-(\bxe - \bxe^2)\left[\langle V_b\rangle \langle W_R^2 \rangle (k)+\tilde{P}_{\delta\delta}(k)\right],
\label{eq:1bubble}
\ee
where 
\be
\tilde{P}_{\delta\delta}(k)=\langle V_b\rangle \int \frac{d^3k'}{(2\pi)^3} \langle W_R^2 \rangle (k')P_{\delta\delta}(|\vec{k}-\vec{k}'|)
\label{eq:ptilde}
\ee
The first term in Eq.~\eqref{eq:1bubble} is the shot noise of the bubbles, which is a direct consequence of randomly placing ionizing galaxies in the Universe. We will later see that this term typically dominates the total correlation function at late times. This can be understood by realizing that the bubbles tend to be larger at late times, assuming that the bubble size increases over time through bubble merging.

Note that when correlating the free electrons with the neutral hydrogen, the one-bubble contribution is always negative, i.e. these are fully anti-correlated when considering just single bubbles in the Universe. 
In the previous section we have also shown that the two-bubble cross-correlation at early times. As reionization proceeds and the neutral hydrogen fraction decreases, the correlation on all scales will become anti-correlated. More importantly, the signal is proportional to the matter power spectrum, which grows as $(1+z)^2$ during matter domination. Within the model applied in this work, we find that at redshift $z<12$, the two-bubble term is negligible compared to the one-bubble term on most scales. 

It was shown by Ref. \citep{Wang:2005my} that $\tilde{P}_{\delta\delta}(k)$ can be approximated as
\be\label{eq:Ptilde}
\tilde{P}_{\delta\delta}(k)&\simeq& \frac{P_{\delta\delta}(k) \langle V_b \rangle\langle\sigma_R^2\rangle}{[(P_{\delta\delta}(k))^2+(\langle V_b \rangle\langle \sigma_R^2\rangle)^2]^{1/2}},
\label{eq:ptildefit}
\ee
which is derived by equating the small and large scale limits of Eq.~\eqref{eq:ptilde} with
\be
\langle\sigma^2_R\rangle&=& \int \frac{k^2 dk}{2\pi^2} \langle W_R^2 \rangle (k) P_{\delta\delta}(k)
\ee
We have found  the  simple fitting solution of Eq.~\eqref{eq:ptildefit} to be accurate to the percent level for most values of $\{\bar{R},\sigma_{\ln R}\}$. In the Appendix we will show that the contribution from $\tilde{P}_{\delta\delta}(k)$ to the one-bubble peak is relatively small for all parameter values in the range of interest for the $\tau-21$ cm cross-correlation, but is non-negligible and relevant at small scales.  

\begin{figure}
\centering
\includegraphics[width=.47\textwidth]{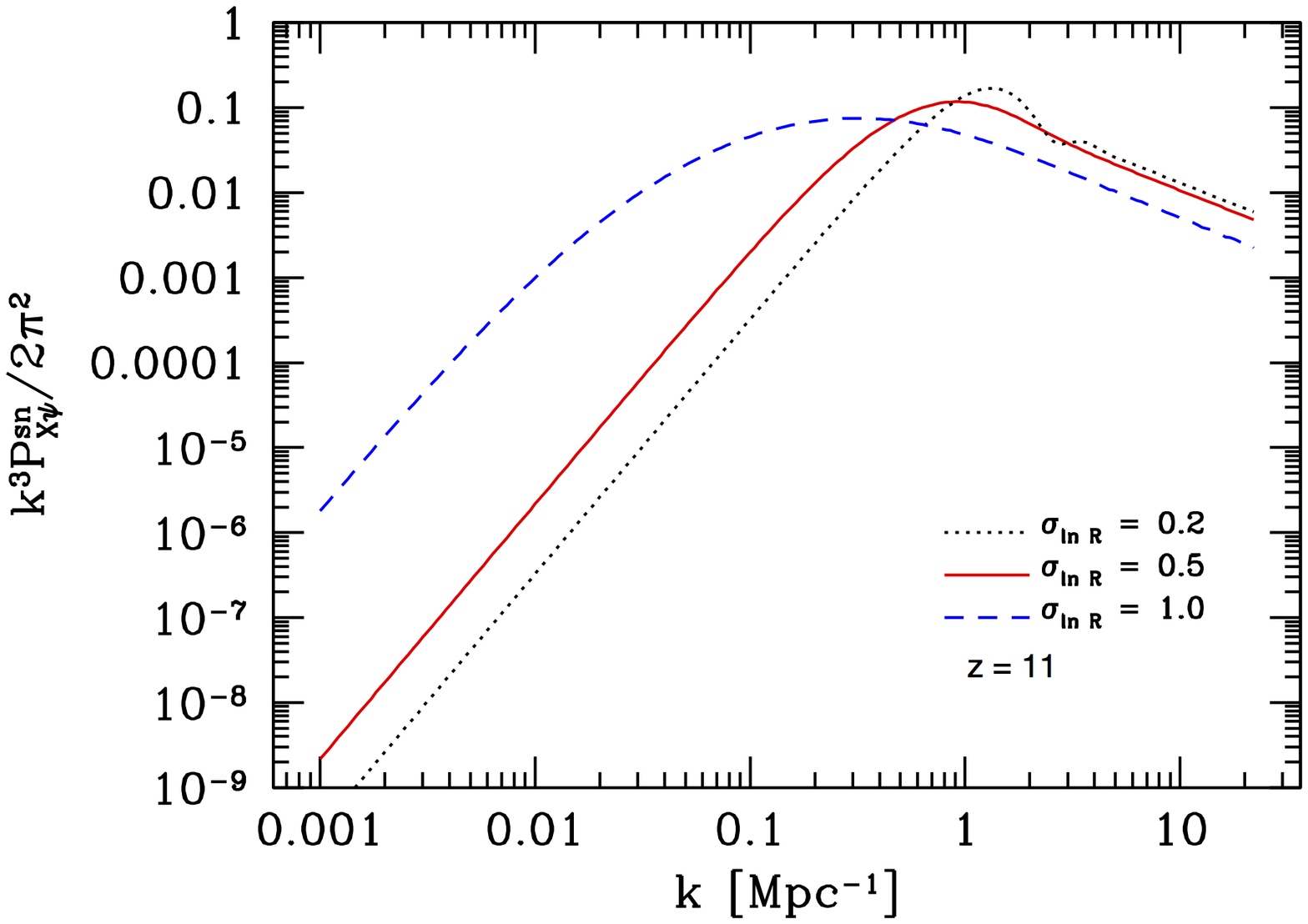}
\caption{The shot noise $P^{sn}_{X\psi}\equiv (\bxe -\bxe^2)\langle V_b\rangle \langle W_R^2\rangle$ for 3 different values of the width of the log-normal distribution. 
}
\label{fig:shotnoise}
\end{figure}

\begin{figure}
\centering
\includegraphics[width=.47\textwidth]{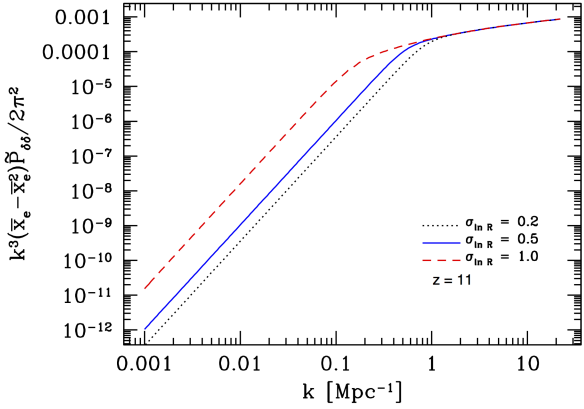}
\caption{ $\tilde{P}_{\delta\delta}$ as defined for 3 different values of the width of the log-normal distribution. This figure shows that this term is only relevant at small scales and does not contribute to the peak of the total correlation function.}
\label{fig:ptilde}
\end{figure}

\begin{figure}
\centering
\includegraphics[width=.47\textwidth]{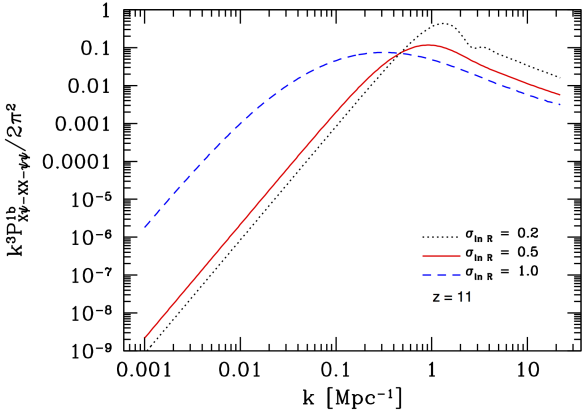}
\caption{ The sum of the previous two figures, the total one-bubble term. The one-bubble term has the same shape for correlations $XX$, $X\psi$ and $\psi\psi$.}
\label{fig:P1bubble}
\end{figure}

We show the shot noise and $\tilde{P}_{\delta\delta}$ in Figs.~\ref{fig:shotnoise} and ~\ref{fig:ptilde}. The sum of these terms gives the one-bubble power spectrum shown in Fig.~\ref{fig:P1bubble}. Note that the one-bubble term from all three possible correlations ($XX$, $\psi\psi$ and $X\psi$) is equivalent up to a sign.

\begin{figure}
\centering
\includegraphics[width=.47\textwidth]{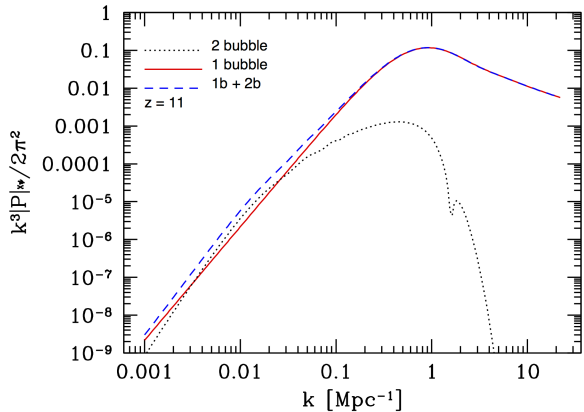}
\caption{$k^3P_{X\psi}(k)/2\pi^2$ with $\sigma_{\ln R} = 0.5$. }
\label{fig:tau21z}
\end{figure}

In the small scale limit,  $\tilde{P}_{\delta\delta}(k)=P_{\delta\delta}(k)$, and the one-bubble term becomes: $P^{1b}_{X\psi}(k) = -(\bxh-\bxh^2) P_{\delta\delta}(k)$, rendering the total cross-correlation negative at these scales (at these scales we are applying a smoothing filter to the two-bubble term, so it effectively does not contribute to the total cross-correlation). 
In Fig. \ref{fig:tau21z} we show the total cross-correlation for $\sigma_{\ln R}=0.5$ at $z=11$ (changing the redshift of the cross-correlation will predominantly affect the average ionization fraction $\bxe$).

\section{Projected cross-correlation}
\label{sec:correlation_2D}

Fourier transforming the dimensionless brightness temperature $\psi$, we can write the spherical harmonic coefficient for the 21 cm fluctuation as
\be
a_{\ell m}^{21} &=& 4\pi (-i)^\ell \int  \frac{d^3k}{(2\pi)^3} \hat{\psi}(\vec{k}) \alpha^{21}_{\ell}(k,z) Y^*_{\ell m}(\hat{k}),
\ee
where
\be
\alpha_{\ell}^{21}(k, z)&=& T_0(z)\int_0^{\infty} d\chi' W_{\chi(z)}(\chi') j_{\ell}(k \chi')
\ee
Note that the response function is centered around $\chi(z) = \chi'$, and in practice we take this distance to be somewhere between $z=0$ and $z=30$.

We can do the same for the optical depth to reionization, i.e. :
\be
\tau_{\ell m} &=&4\pi (-i)^{\ell} \int \frac{d^3k}{(2\pi)^3} X(\vec{k}) \alpha^{\tau}_{\ell}(k)Y^*_{\ell m}(\hat{k}), \nonumber\\
\ee
with
\be
\alpha^{\tau}_{\ell}(k) = (1-Y_p)\sigma_T \frac{\rho_{b,0}}{m_p}  \int_0^{\chi_*}\frac{d\chi' }{a^2} j_{\ell}(k \chi'),
\ee
where $\chi_*$ corresponds to the distance to last scattering. 

Cross-correlating the two maps yields
\be
\langle \tau_{\ell m} a^{21*}_{\ell'm'}(z) \rangle &=& \delta_{\ell\ell'}\delta_{mm'}C^{\tau,21}_{\ell}(z)\nonumber\\
&=& \int \frac{dk}{k} \Delta^2_{X \psi}(k) \alpha^{\tau}_{\ell}(k) \alpha_{\ell}^{21}(k,z)\nonumber \\
\label{eq:Clclean}
\ee
Here, $\Delta^2_{X \psi}= k^3 P_{X\psi}/(2\pi^2)$. 

Let us consider the cross-correlation in the Limber approximation. Under this approximation, we can assume that the Bessel functions are small, $j_{\ell}(x)\ll1$, for $x< \ell$ and peak when $x\sim \ell$. The integral over comoving momentum $k$ will get most of its contribution from modes $k\sim \ell/\chi$. Therefore we can make the approximation that $\Delta^2_{X\psi}(k)\sim \Delta^2_{X\psi}(\ell/ \chi)$ and re-write Eq.~\eqref{eq:Clclean} as
\be
C_{\ell}^{\tau,21}(z)&=&(1-Y_p)\frac{T_0(z) \rho_{b,0}\sigma_T}{m_p}\nonumber \\
&&  \int_0^{z_*} \frac{dz_1'}{H(z_1')} (1+z_1')^2 \int_0^{\infty} \frac{dz_1'}{H(z_1')}   W_{z}(\chi(z_1'))\nonumber\\
&& \times 4\pi \int \frac{dk}{k} \Delta^2_{X\psi}(k) j_{\ell}(k \chi(z_1')) j_{\ell}(k \chi(z_2'))\nonumber \\
\label{eq:Clraw}
\ee 
Again, in practice we take the window function to be centered around $0\le z \le 30$.
 
In the Limber approximation we can perform the $k$ integral over the product of Bessel functions as:
\be
\int_0^{\infty} dk k^2 j_{\ell}(k  \chi(z_1)) j_{\ell}(k \chi(z_2))=\frac{\pi}{2} \frac{\delta( \chi(z_1)- \chi(z_2))}{ \chi ^2}
\ee 

Thus, we can write the angular cross spectrum as
\be
C_{\ell}^{\tau,21}(z)&=&(1-Y_p)\frac{T_0(z) \rho_{b,0}\sigma_T}{m_p}\int_0^{\infty} dz'  \frac{W_{z}(\chi(z'))}{H^2} \nonumber \\
&&  \times \left| \frac{d\chi}{dz} \right|^{-1}  \left(\frac{1+z'}{ \chi(z')}\right)^2P_{X\psi}\left(\frac{\ell}{\chi(z')},z'\right)
\ee

We will assume that the window function is a Gaussian centered around  redshift $z$ with width $\delta \chi$ given by
\be
\delta \chi \simeq \left(\frac{\Delta \nu}{0.1 \mathrm{MHz}} \right)\left( \frac{1+z}{10} \right)^{1/2} \left(\frac{\Omega_m h^2}{0.15}\right)^{-1/2} \mathrm{Mpc}
\ee 
where $\Delta \nu$ is the bandwidth frequency of the instrument. We have taken into account that the power spectrum explicitly depends on redshift. 

\begin{figure}
\centering
\includegraphics[width=.47\textwidth]{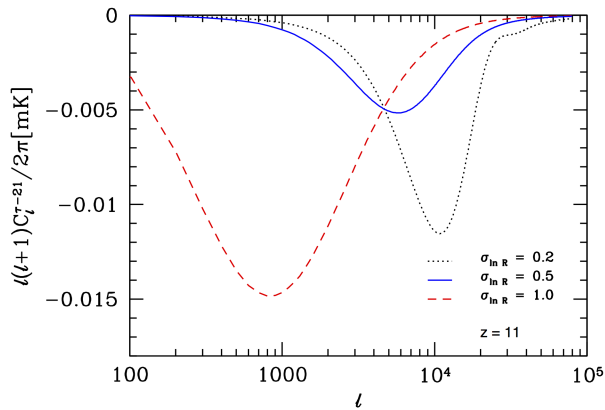}
\caption{The cross-correlation between 21 cm temperature brightness fluctuations and the optical depth $\tau$ at $z=11$. Here we used a Gaussian window function with bandwidth frequency $\Delta \nu = 0.2$ MHz.}
\label{fig:cltau21}
\end{figure}

We show the angular cross-correlation for $\Delta \nu = 0.2$ MHz and several values of $\sigma_{\ln R}$ in Fig.~\ref{fig:cltau21}.

In the previous sections we have shown that the cross-correlation between free electrons and neutral hydrogen has a strong dependence on the parameters that determine the bubble distribution as well as its bias. The location of the peak is set by an effective scale, which we derive in the Appendix. We will see that for a log-normal bubble distribution, this effective scale is exponential in the width of the distribution  and inversely proportional to the average bubble radius. Therefore, a small change in the width of the distribution can produce a large change in the location of the peak. We have also shown that at early times the two-bubble term can be positively correlated at large scales. The positive contribution to the correlation function at large scales eventually vanishes when the universe further reionizes. However, if the bubbles are small enough, a positive contribution to the correlation function could persist until late times. Oppositely, if bubbles are relatively large (a few Mpc), the shot noise, which is negative for all scales, will dominate the correlation function. For the reionization parameterization in this paper, the shot noise is the dominated term at the most relevant redshifts (peaking around $\bxe = 0.5$).

\section{Is the cross-correlation detectable?}
\subsection{Signal-to-noise}
\label{sec:Signal_to_Noise}
In this section we will determine if the cross-correlation is detectable. 
An important issue that we will address here are the foregrounds. As previously mentioned, the $21$ cm emission should be swamped by foregrounds, dominated on large scales by polarized Galactic synchrotron, with a total intensity of 3-4 orders of magnitude larger than the 21 cm brightness from reionization. On small scales, the redshifted 21 cm brightness is obscured by extragalactic sources \citep{Shaver:1999gb,2008MNRAS.389.1319J}. 
We do not know the spectral dependence of all these foregrounds, but in general we can assume that they are relatively smooth in frequency along the line of sight, as they are associated with same source (e.g. our own Galaxy). In principle, one can therefore remove a large part of the (large scale) foregrounds by removing the largest modes along the line of sight (see e.g. \cite{Liu:2011ih} for a recent discussion).

However, when cross-correlating the 21 cm field with the optical depth, we want to keep the largest modes, to which the integrated optical depth is most sensitive. Hence, we will  keep the foregrounds in the observed maps and show that the cross-correlation between foregrounds in the 21 cm field and in the CMB should be small.
In order to neglect the cross-correlation of the foregrounds between $\tau$ and $21$ cm, we typically need the foreground of the CMB  to be $\leq10^{-5}$ times the signal \citep{Liu:2009qga} (given that $a_{\ell m}^{f,21} \sim 10^5 a_{\ell m}^{21}$). 
The synchrotron emission is roughly equal to the CMB signal at $1$ GHz. Therefore, if we assume that the synchrotron scales as $\nu^{-3}$ \citep{Kogut:2007tq}, at $94$ GHz ($W$ band) we have $a_{\ell m}^{synchrotron}\sim \times10^{-6} a_{\ell m}^{CMB}$. Thus, we estimate that the signal will be larger than the remaining foregrounds after cross-correlating the two maps. 

Additionally, by not removing the foregrounds, the 21cm foregrounds will effectively act as noise term in the cross-correlation. In other words, even in the absence of correlation between foregrounds, there is still a finite probability that any given data point in the $\tau$ map will correlate with a foreground measurement from 21 cm, i.e. the induced noise contains a term $\langle \tau_{\ell m} a_{\ell'm'}^{f,21}\rangle$, where the latter is the spherical harmonic coefficient of the 21 cm foreground map. 

Unfortunately, we do not know exactly what the level of synchrotron foreground is, but typically $C^f_{\ell}\sim k \ell^{-\alpha}$, with $2< \alpha <4$. We will assume that the synchrotron emission scales as $\nu^{-3}$. \cite{LaPorta:2008ag} showed synchrotron emission at 480 MHz has a normalized amplitude of $100 \; \mathrm{mK^2}  < C^f_{\ell=100} <10000 \; \mathrm{mK^2}$, with the actual amplitude and slope depending on the position in the sky. 

Although we cannot remove the foregrounds through implementing a large scale cutoff, we can alternatively try to remove a substantial part of galactic foreground emission.
If there is a large correlation between different frequencies of the foreground maps, one could measure the foregrounds at high frequency (corresponding to a completely ionized universe and, hence, with no signal in the cross-correlation), extrapolate with an appropriate scaling $\sim \nu^{-3}$ and subtract those from the high redshift maps \footnote{ This is scaling is approximate and simplistic. In reality one would probably have to consider a slope that changes as a function of scale and frequency.  We are assuming the scaling will be further understood as a function of frequency once we are capable of performing this cross correlation.} \citep{Shaver:1999gb}. If the correlation between different maps at high frequencies is of order $0.9-0.99$, one could reduce the overall amplitude of the foreground by a factor of $10-100$ and the power by a factor of a $100-10^4$ \footnote{Note that this is a very crude estimate. For example, it might be relevant to consider 21 cm signals after reionization (at low $z$)
due to residual neutral hydrogen, primarily in Damped Ly$\alpha$ absorbers (DLA's). }. In addition, \cite{Liu:2012xy} showed that down weighting the most heavily contaminated regions in the sky can reduce the effective foreground as much as a factor of $2$. Note that this approach is different from the usual spectral fitting techniques \citep{Shaver:1999gb,Santos:2004ju,Wang:2005zj,McQuinn:2005hk,2009MNRAS.397.1138H,2010MNRAS.402.2279J,2012MNRAS.423.2518C,2013MNRAS.429..165C}.

At small scales we expect extra galactic radio sources to dominate the foregrounds. However, there are several strategies that will likely suppress the noise term due to correlations between millimeter and radio emission: Ê
(i) since bright sources are expected to dominate the variance at radio frequency \citep{2010A&ARv..18....1D,2011A&A...536A..13P,2013MNRAS.432.2625H}, Êthese sources
can be masked at ~5 $\sigma$. ÊThis will suppress the radio source contribution without removing very much of the sky;Ê(ii) at a given location in the map,
the radio data short wards and long-wards of the 21 cm radio emission can be used to remove both galactic and extragalactic foreground by assuming
that the sources at a given location can be fit by a power lawÊ; (iii) at millimeter and sub millimeter wavelengths,
multi-frequency data can be used to separate CMB signal from dusty galaxy foregrounds.ÊThe Planck data shows that the 353 GHz data can be used
to remove $>90\%$ of the dusty galaxy foreground at 220 GHz \citep{2013arXiv1303.5073P}. All of these strategies will likely be employed to remove foregrounds in both maps.

We will consider a case in which the angular power spectrum of the foreground is given by:
\be\label{eq:Clf}
C_{\ell}^f(z) \simeq {10^6\over 2}\frac{1}{c_f}\; \mathrm{mK}^2 \ell^{-3}\left({f(z)\over 480 \; \mathrm{MHz}}\right)^{-3},
\ee
where $f(z)$ corresponds to the frequency of the redshift considered and $100\leq c_f \leq 10^4$ is the foreground reduction factor that we can hope to achieve through a measurement at low redshift. As we will show later, the signal-to-noise of the cross-correlation does not vary substantially for different values of $c_f$ in this range.

We will assume a noise power spectrum given by \citep{Morales:2004ca,McQuinn:2005hk,Adshead:2007ij,Mao:2008ug},
\be
N_{\ell}^{21,21} &=& \frac{2\pi}{\ell^2}(20 \; \mathrm{mK})^2 \left[\frac{10^4 \mathrm{m^2}}{A_{\mathrm{eff}}}\right]^2 \left[\frac{10'}{\Delta \Theta}\right]^4\left[\frac{1+z}{10}\right]^{9.2} \nonumber\\
&& \times \left[ \frac{\mathrm{MHz}}{\Delta \nu}\frac{100\;\mathrm{hr}}{t_{\mathrm{int}}}\right]
\ee
We will do forecasts for a total integration time of 1000 hours, and a beam with an angular diameter of $\Delta \Theta = 9$ arcmin. We set the bandwidth to $\Delta \nu = 0.2$ MHz.
For a LOFAR \citep{2013A&A...556A...2V} type experiment, we use $A_{\mathrm{eff}} = 10^4$ m$^2$ and for a Square Kilometer Array (SKA) type experiment we use $A_{\mathrm{eff}} = 10^5$ m$^2$.

On the CMB side, experiments are rapidly improving \citep{Planck:2006aa, Niemack:2010wz, Austermann:2012ga, Zaldarriaga:2008ap}, with high sensitivity experiments coming soon (Planck, ACTPol, SPTPol, CMBPol) and we should have observations of the E- and B-modes polarization spectra up to small scales in the near future. We will now consider a next generation polarization experiment that allows us to reconstruct a map of the optical depth $\tau_{\ell m}$ with the estimator proposed by \cite{Dvorkin:2008tf}.
This estimator was built to extract the inhomogeneous reionization signal from future high-sensitivity measurements of the cosmic microwave background temperature and polarization fields. 
\cite{Dvorkin:2008tf} wrote a minimum variance quadratic estimator for the modes of the optical depth field given by:

\be\label{eq:tau_estimator}
\hat{\tau}_{\ell m}&=& N^{\tau\tau}_{\ell} \sum_{\ell_1m_1\ell_2m_2} \Gamma_{\ell_1\ell_2\ell}^{EB}\left(\begin{array}{ccc} \ell_{1} & \ell_{2} & \ell \\ m_{1} & m_{2} & m \end{array}\right)\nonumber \\
&& \times  \frac{a^{E_*}_{\ell_1m_1}a^{B_*}_{\ell_2m_2}}{(C_{\ell_1}^{EE}+N_{\ell_1}^{EE})(C_{\ell_2}^{BB}+N_{\ell_2}^{BB})},
\ee
where $C_\ell^{EE}$ and $C_\ell^{BB}$ are the $E$- and $B$-mode polarization power spectra. $N_{\ell}^{EE}$ and $N_{\ell}^{BB}$ correspond to the CMB noise power spectra, and are given by:
\be
N_{\ell}^{EE}=N_{\ell}^{BB}=\Delta_P^2 \exp\left({\ell(\ell+1)\theta^2_{\rm FWHM}\over 8\ln(2)}\right),
\ee
where $\Delta_P$ is the detector noise and $\theta_{\rm FWHM}$ is the beam size.

The coupling $\Gamma_{\ell_1\ell_2\ell}^{EB}$ can be written as
\be
\Gamma_{\ell_1\ell_2\ell}^{EB}&=&\frac{C_{\ell_1}^{E_0E_1}}{2i}  \sqrt{\frac{(2\ell_{1}+1)(2\ell_{2}+1)(2\ell+1)}{4\pi}}\nonumber \\
&&\times \left[\left(\begin{array}{ccc} \ell_{1} & \ell_{2} & \ell \\ -2 & 2 & 0 \end{array}\right)-\left(\begin{array}{ccc} \ell_{1} & \ell_{2} & \ell \\ 2 & -2 & 0 \end{array}\right)   \right]
\ee 
Here $C_{\ell_1}^{E_0E_1}$ is the cross-power spectrum between the CMB $E_0$-mode polarization without patchy reionization and the response field to $\tau$ fluctuations $E_1$. ($C_\ell^{E_0E_1}$ is positive at large scales due to Thomson scattering, and negative at small scales due to the screening).

Furthermore, the reconstruction noise power spectrum is given by:
\be 
 N^{\tau\tau}_{\ell}&=&\left[\frac{1}{2\ell+1}\sum_{\ell_1\ell_2} \frac{|\Gamma_{\ell_1\ell_2\ell}^{EB}|^2}{(C_{\ell_1}^{EE}+N_{\ell_1}^{EE})(C_{\ell_2}^{BB}+N_{\ell_2}^{BB})}\right]^{-1} \nonumber \\
 \label{eq:taunoise}
\ee 

We note that the main source of contamination in reconstructing $\tau(\hat{n})$ comes from the non-Gaussian signal from gravitational lensing of the CMB. In principle, unbiased estimators that simultaneously reconstruct the inhomogeneous reionization signal and the gravitational potential can be constructed \citep{Su:2011ff}. For purposes of simplicity, we will estimate our cross-correlation using the estimator given by Eq. \eqref{eq:tau_estimator}, but the results in this work are straightforward to generalize.

Given that the $\tau$ map is not sensitive to redshift  (\cite{Dvorkin:2008tf} showed that the estimator is only sensitive to one principal component in redshift), while the 21 cm map can be reconstructed on redshift slices, we will give a weight to the cross-correlation. This weight will be built in order to maximize the signal to noise, in the same spirit as in \cite{Peiris:2000kb}, where a weight was derived for the cross-correlations between CMB and Galaxy surveys. We write the weighted 21 cm maps as
\be
\tilde{a}_{\ell m}^{21}(z)=\int_0^z dz' a_{\ell m}^{21}(z')w_\ell(z')
\ee
We then want to maximize
\be
\chi_{\ell m}^2 = \frac{\langle\tau_{\ell m}^*\tilde{a}_{\ell m}^{21}(z)\rangle^2}{\langle \tau_{\ell m}\tau_{\ell m}^*\rangle \langle \tilde{a}_{\ell m}^{21}(z)\tilde{a}_{\ell m}^{21 *}(z)\rangle},
\ee 
and, in doing so, we find:
\be
w_\ell(z) &=& \frac{C_\ell^{\tau,21}(z)}{(C_\ell^{21,21}+N_\ell^{21,21}+C_\ell^f)(z)},
\ee
which is nothing else then the projected signal over the projected noise. Note that we have included the foreground as a source of noise as explained before. 

We can now compute  the signal to noise for the $\tau$-$21$ cm cross-correlation as
\be
\left(S\over N\right)^2& =&  f_{\rm sky}  \sum_\ell (2\ell+1) \times \nonumber \\
&&\int dz {|C_\ell^{\tau,21}(z)|^2 \over (C_\ell^{\tau,\tau}+N_\ell^{\tau\tau}) (C_\ell^{21,21}+N_\ell^{21,21}+C_\ell^f)(z)} \nonumber \\
\ee

\begin{figure}
\centering
\includegraphics[width=.47\textwidth]{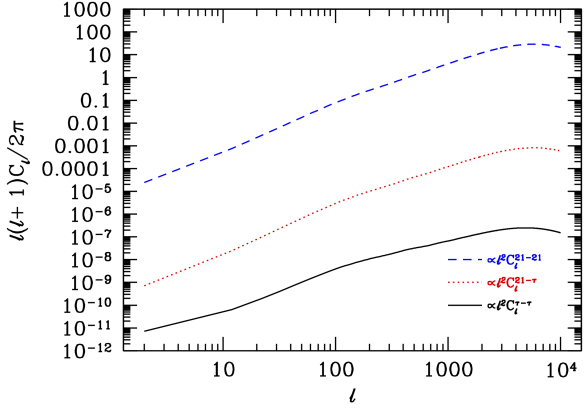}
\caption{The different power spectra used to compute the signal to noise in Fig. \ref{fig:StoN_tau21_LOFAR_SKA}. The sign of $C_{\ell}^{21-\tau}$ has been inverted for the sake of comparison. The spectra are shown at $z=11$. The bubble radii as a function of redshift, the bubble bias and the bubble number density are all determined through our toy reionization model of Eq.~\protect{\eqref{eq:barnb}} as explained in section  \S\ref{sec:reionization_model}.The noise to the $\tau$ estimator is given by Eq.~\protect{\eqref{eq:taunoise}} \citep{Dvorkin:2008tf}. }
\label{fig:cls}
\end{figure}
 
\begin{figure}
\centering
\includegraphics[width=.47\textwidth]{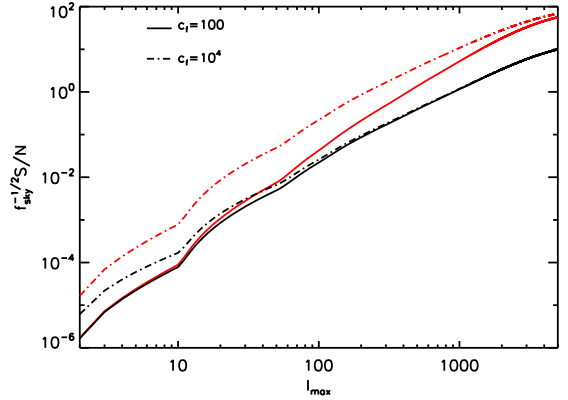}
\caption{Total signal-to-noise for the $21$ cm-$\tau$ cross-correlation as a function of $\ell_{max}$ for a model with a log normal bubble distribution with a width  $\sigma_{\ln R}=0.5$. We consider an experiment with CMB noise power spectra given by $\Delta_P=0.3 \mu$K-arcmin and beam size $\Theta_{FWHM}=1$ arcmin. The foreground angular power spectrum is given by Eq.~\eqref{eq:Clf}. We show forecasts for an experiment with the planned noise level of SKA (in red lines) and the planned noise level of LOFAR (in black lines). Note that the y-axis represents the product of the signal-to-noise and the fraction of the sky covered. For an experiment like LOFAR (with  $f_{\rm sky}=0.001$), we find $S/N=0.2$ and for SKA (with $f_{\rm sky}=0.25$), we get $S/N=16$ at $\ell_{\rm max}=3000$ with $c_f=100$. At small scales does not vary substantially when considering different values of the parameter $c_f$, which represents the level of foreground subtraction.}
\label{fig:StoN_tau21_LOFAR_SKA}
\end{figure}

In Fig. \ref{fig:StoN_tau21_LOFAR_SKA} we assess the level of detectability for a reionization history with $\sigma_{\ln R}=0.5$ (see Fig.~\ref{fig:cls} for an example angular spectra at $z=11$). Again, to generate the spectra we use a Gaussian window function with $\Delta \nu = 0.2$ MHz. 
We find that the signal to noise reaches $f_{sky}^{-1/2}S/N=0.2$ at $\ell_{\rm max}=3000$ for a LOFAR type experiment (with $f_{\rm sky}=0.001$), while for a SKA type experiment (with $f_{\rm sky}=0.25$) it reaches $f_{sky}^{-1/2}S/N=16$, with $f_{\rm sky}$ being the fraction of the sky covered. 
We note that at small scales, the signal to noise does not vary substantially when considering different values of the parameter $c_f$, that represents the level of foreground subtraction.

\subsection{Reionization parameters}

In this section we will assess what we can learn about reionization by studying the $21$ cm-$\tau$ cross-correlation.
We will forecast parameter uncertainties in the following parameters $\pi=\{\tau,\Delta y\}$.

As a forecasting tool we will use a Fisher matrix analysis, where the Fisher matrix is given by \citep{Tegmark:1996bz}:
\be
F_{\mu\nu}&=&f_{sky}\sum_{\ell}(2\ell+1)\times\nonumber\\
&&\int dz{(\partial C_{\ell}^{\tau,21}(z)/\partial \pi_\mu) (\partial C_{\ell}^{\tau,21}(z)/\partial \pi_\nu)\over (C_\ell^{\tau,\tau}+N_\ell^{\tau\tau}) (C_\ell^{21,21}+N_\ell^{21,21} + C_\ell^f)(z)} \nonumber \\
\ee
where $\mu$ and $\nu$ run over the parameter modes.

The rms uncertainty on the parameter $\pi_\mu$ is given by $\sigma(\pi_\mu)=(F^{-1}_{\mu\mu})^{1/2}$ if the other parameters are marginalized. If the remaining parameters are assumed fixed, then the rms is $\sigma(\pi_\mu)=(F_{\mu\mu})^{-1/2}$.

We consider a next generation polarization experiment with noise power spectrum given by $\Delta_P=0.3 \mu$K-arcmin and beam size $\Theta_{FWHM}=1$ arcmin.

When assuming an experiment with the planned noise level of SKA and $f_{\rm sky}=0.25$, the width of reionization $\Delta y$ can be constrained at the $10\%$ level, and $\tau$ at the $4\%$ level, when the remaining parameters are considered fixed.

\begin{figure}
\centering
\includegraphics[width=.47\textwidth]{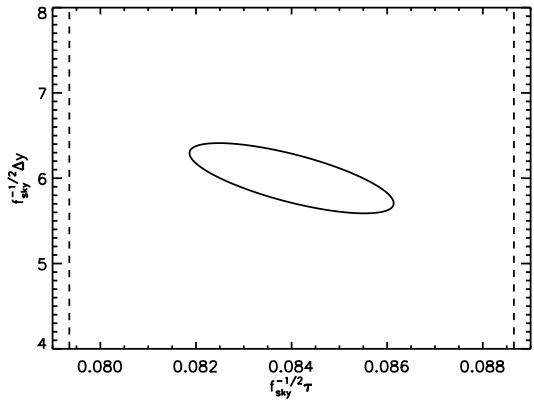}
\caption{Forecasted uncertainties on the width of reionization parameter $\Delta y$ and the optical depth $\tau$ (assuming that the other parameters are fixed) for an experiment with the planned noise level of SKA. Note that the y-axis represents the product of the parameter value and the fraction of the sky observed. For reference, the dashed lines correspond to the Planck priors on the optical depth.}
\label{fig:error_ellipse_tau_Deltay}
\end{figure}

We show the error ellipses for the optical depth and the width of reionization in Fig. \ref{fig:error_ellipse_tau_Deltay}.
The Planck priors are shown in dashed lines. 

\section{Discussion and Conclusion}
\label{sec:Discussion and Conclusion}

We investigated the correlation between free electrons, traced by the optical depth $\tau$, and neutral hydrogen, traced through the emission of 21 cm photons, during the epoch of reionization. To compute the cross-correlation we used a simple model where patches of ionized gas are represented by spherical bubbles. The cross-correlation will depend on the presence or absence of these bubbles (the one-bubble term) and the clustering of bubbles (the two-bubble term). As expected, the cross-correlation is negative on small scales, where it is dominated by the shot noise of the bubbles. On large scales, the two-bubble term can render the correlation positive as long as the effective bias $b_{\mathrm{eff}}$ is large or $\bxe$ is small. Small bubbles at a fixed neutral hydrogen fraction imply a small bubble bias, hence the two-bubble term has a suppressed (positive) amplitude. A larger correlation could be driven by the ionization fraction, but within a bubble merger scenario the smallest bubbles are expected at early times, when the ionization fraction is small and the total matter power spectrum is suppressed. 

The anti-correlation peak, set by the sum of the one and the two-bubble terms, depends critically on the distribution of the ionized bubbles. Consequently, a measurement of the cross-correlation allows us to probe the parameters relevant for the reionization history. In principle a measurement of a positive correlation at early times, would theoretically allow us to entangle the degeneracy between $\tau$ and the bubble bias. However, we showed that the two-bubble term typically has a very small amplitude. 

One major obstacle in measuring the 21 cm emission from the EoR are the large foregrounds at these frequencies. For the auto-correlation, any detection requires a careful removal of foregrounds, which typically results in the removal of the largest modes along the line of sight. The advantage of the cross-correlation is that foregrounds in the measurement of $\tau$ are weakly correlated with those in the $21$ cm field. Therefore,  the cross-correlation is less sensitive to the detailed understanding of the foregrounds. 

In this paper we have computed the signal to noise of the cross-correlation using the estimator for inhomogeneous reionization  $\hat{\tau}_{\ell m}$ proposed by \cite{Dvorkin:2008tf}. In our computation there is very little contribution from any positive correlation at large scales coming from two-bubble term, and most of the signal comes from the shot noise. Because a measurement of the optical depth gets most of its signal from the long wavelength mode along the line of sight, we left the 21 cm foregrounds as a noise term. Although the signal to noise per mode is small, the large number of modes allows for a detection when considering a next generation 21 cm experiment cross-correlated with a CMB experiment that measures the polarization $B$-modes in most of the sky. We expect that around the time SKA observes a large part of the sky, CMB experiments will have improved to the level that we are able to reconstruct a map of $\tau_{\ell m}$. 
Although the auto-correlation of both maps will give significant insight into reionization, cross-correlating these maps will provide us with a complementary probe. 
We find that a measurement of this cross-correlation with a detector noise level of SKA (and $f_{\rm sky}=0.25$) on the 21cm side and noise level of a next generation polarization type experiment on the CMB side constrains the width of the ionization history at the $10\%$ level and the optical depth at the $4\%$ level.

\acknowledgements
The authors would like to thank Renyue Cen, Enrico Pajer, Fabian Schmidt, Kendrick Smith, and Matias Zaldarriaga for useful discussions. P.D.M. is supported by the Netherlands Organization for Scientific Research (NWO), through a Rubicon fellowship.
C.D. is supported by the National Science Foundation grant number AST-0807444, NSF grant number PHY-0855425, and the Raymond and Beverly Sackler Funds. P.D.M. and D.N.S. are in part funded by the John Templeton Foundation grant number 37426. 

\appendix
\section{Reionization model dependence}

The choice of a log-normal distribution is motivated in part by simulations in \citep{Zahn:2006sg} and \citep{Wang:2005my}. In this appendix we derive constraints on the relative contributions of the various terms of the cross correlation as a function $b$, $\bar{R}$ and $\sigma_{\ln R}$. The aim of this appendix is to show that in most realistic scenarios, the shot noise is generally the dominating term, independent of reionization details. 

\subsection{Log-normal distribution}

We first start by investigating the implications of a log-normal distribution for the bubble radius.

The bubble distribution is given by
\be
P(R, \sigma_{\ln R}) = \frac{1}{R} \frac{1}{\sqrt{2\pi\sigma^2_{\ln R}}} e^{-[\ln (R/\bar{R})]^2/(2\sigma_{\ln R}^2)}
\ee
Given this distribution we can compute the average bubble size:
\be
\langle V_b \rangle &=&\int dR P(R) V_b(R) = \frac{4\pi \bar{R}^3 }{3} e^{9\sigma_{\ln R}^2/2}
\ee
To address the dependence of the resulting correlation function, we also need the variance 
\be
\langle V_b^2 \rangle &=&\int dR P(R) V^2_b(R) = \frac{(4\pi)^2 \bar{R}^6 }{9} e^{18\sigma_{\ln R}^2}
\ee
The amplitude of the one and two-bubble terms, and the relevant scale where these peak, strongly depend on the window function
\be
W_R(k)& = &\frac{3}{(kR)^3}\left[\sin (kR)-kR\cos(kR)\right] 
\ee

\begin{figure}
\centering
\includegraphics[width=.47\textwidth]{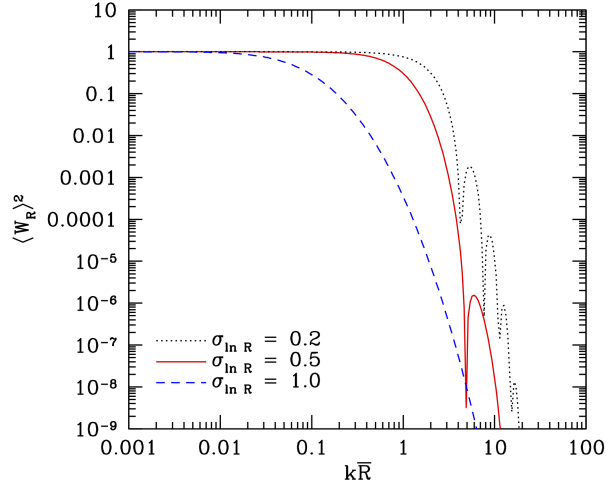}
\caption{$\langle W_R\rangle^2$ for different values of $\sigma_{\ln R}$.}
\label{fig:windowav}
\end{figure}

\begin{figure}
\centering
\includegraphics[width=.47\textwidth]{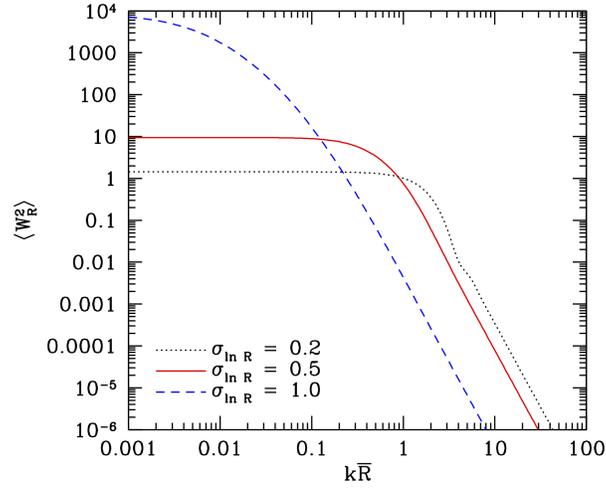}
\caption{$\langle W_R^2\rangle$ for different values of $\sigma_{\ln R}$. Note that the maximum amplitude in the limit $k\bar{R}\ll1$ grows exponentially in $\sigma_{\ln R}$, which will be relevant for the relative contribution of the shot noise with respect to the one and two-bubble terms.}
\label{fig:windowvar}
\end{figure}

Recall the the volume averaged window function and window function squared (shown in Figs. \ref{fig:windowav} and \ref{fig:windowvar}) are defined as 
\be
\langle W_R\rangle(k)&=& \frac{1}{\langle V_b\rangle} \int_0^{\infty} dR P(R) V_b(R) W_R(k R),
\ee
and 
\be
\langle W_R^2\rangle(k)&=& \frac{1}{\langle V_b\rangle^2} \int_0^{\infty} dR P(R)V_b^2(R) W_R^2(k R)
\ee

The total correlation function can be written as 
\be
P_{X\psi} = P^{1b}_{X\psi} + P^{2b}_{X\psi},
\ee
where the one-bubble contribution consists of two relevant terms: the shot noise and the power spectrum $\tilde{P}_{\delta\delta}$ given by Eq.~\eqref{eq:Ptilde}. 

Let us define the following relevant ratios:
\be
R^{2b-sn}\equiv P^{2b}_{X\psi}/P_{X\psi}^{sn}
\ee
and
\be
R^{1b-sn}\equiv \tilde{P}_{\delta\delta}/P^{sn}_{X\psi},
\ee
where $P^{sn}_{X\psi}$ is the contribution from the shot noise, which is given by 
\be
P^{sn}_{X\psi}&=&-(\bxe - \bxe^2)\langle V_b\rangle \langle W_R^2 \rangle (k)
\ee

In the limit of small comoving momenta (large scales), $\langle W_R\rangle\rightarrow 1$, and $\langle W_R^2\rangle\rightarrow  \langle V_b^2 \rangle / \langle V_b\rangle ^2$.

For small scales, we have:
\be
\langle W_R^2\rangle(k, \bar{R}, \sigma_{\ln R}) &\sim& \frac{9}{2k^4 \langle V_b\rangle^2} \int_0^{\infty} \frac{dR}{R^4} P(R) V_b^2(R)\nonumber \\
&=& \frac{9 }{2 k^4\bar{R}^4}  e^{-7\sigma_{\ln R}^2}
\ee
Roughly speaking, we know that the contribution from the shot noise term will be constant and have a peak at some characteristic scale after which it will decrease as $\propto 1/k^4$. Furthermore, up until that characteristic scale, the amplitude of the shot noise is boosted with respect to all the other terms as $ \langle V_b^2 \rangle / \langle V_b\rangle ^2 = e^{9\sigma_{\ln R}^2}$.


The characteristic scale is determined by equating the two limiting cases \citep{Mortonson:2006re}, i.e. 
\be
 \frac{(4\pi)^2 \bar{R}^6 }{9} e^{18\sigma_{\ln R}^2} =  \frac{(4 \pi)^2 \bar{R}^2}{2k^4}  e^{2\sigma_{\ln R}^2},
\ee
This  tells us that the shot noise roughly peaks around
\be\label{eq:kpeak}
k_{peak} = \left(\frac{9}{2 \bar{R}^4} e^{-16\sigma_{\ln R}^2}\right)^{1/4}
\ee

The second contribution to the one-bubble term comes from $\tilde{P}_{\delta\delta}(k)$, which is given by
\be
\tilde{P}_{\delta\delta}(k)=\langle V_b\rangle \int \frac{d^3k'}{(2\pi)^3} \langle W_R^2 \rangle (k')P_{\delta\delta}(|\vec{k}-\vec{k'}|)
\ee

\begin{figure}
\centering
\includegraphics[width=.47\textwidth]{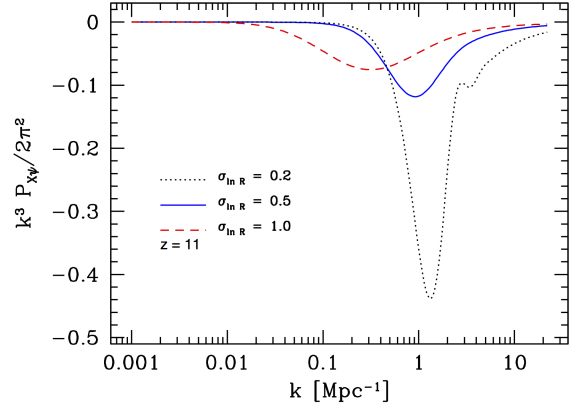}
\caption{$k^3P_{X\psi}(k)/2\pi^2$ for $\bar{R}=1$ Mpc, $b=6$ at different values of  $\sigma_{\ln R}$.}
\label{fig:ptotalXpvar}
\end{figure}

As it was shown by \cite{Wang:2005my}, in the large scale limit we have: 
\be
\lim_{k \bar{R}\ll 1} \tilde{P}_{\delta\delta}(k) \simeq \langle V_b\rangle  \int _0^{\infty} \frac{dk }{2\pi^2} k^2 \langle W_R^2\rangle (k ) P_{\delta\delta}(k)
\ee
In this limit, $\langle W_R^2\rangle \rightarrow \langle V_b^2\rangle/ \langle V_b\rangle^2$, which allows us to put the following constraint on the amplitude of $\tilde{P}_{\delta\delta}(k)$
\be
\tilde{P}_{\delta\delta} \lesssim \frac{\langle V_b^2\rangle}{ \langle V_b\rangle} \int_0^{k_{peak}} k^2 dk P_{\delta\delta}(k)
\ee
The relative peak amplitude between the two contributions in the one-bubble term is therefore given by 
\be
R^{1b-sn}\equiv \tilde{P}_{\delta\delta}/P^{sn}_{X\psi} \lesssim  \frac{1}{2\pi^2} \int_0^{k_{peak}} k^2 dk P_{\delta\delta}(k)
\ee
Since $k_{peak}$ depends on the bubble radius and on $\sigma_{\ln R}$, so does the relative contribution. Generally speaking, a narrower distribution (with smaller $R_0$) leads to a larger contribution from $\tilde{P}_{\delta\delta}$ to the total one-bubble term. That being said, even for very narrow distributions and very small average bubble radius we find that the total contribution to the peak does not exceed more then a few percent, i.e., in realistic scenarios the shot noise term dominates the total one-bubble term  (see e.g. \cite{Zahn:2010yw}). In Fig. \ref{fig:ptotalXpvar}  we plot the cross-correlation $X\psi$ for different values of $\sigma_{\ln R}$, confirming our estimate for the peak sale in Eq. \eqref{eq:kpeak}. 

\begin{figure}
\centering
\includegraphics[width=.46\textwidth]{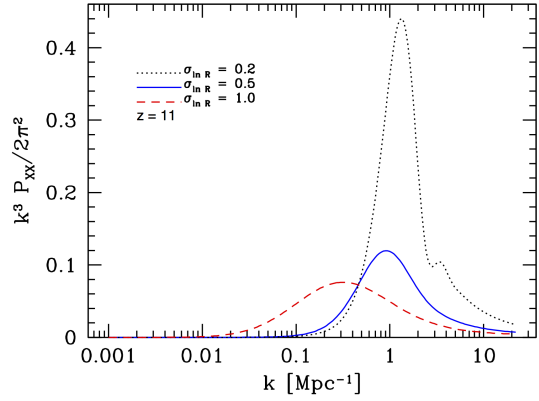}
\caption{$k^3P_{XX}(k)/2\pi^2$ for $\bar{R}=1$ Mpc, $b=6$ at different values of  $\sigma_{\ln R}$.}
\label{fig:ptotalXXvar}
\end{figure}

\begin{figure}
\centering
\includegraphics[width=.47\textwidth]{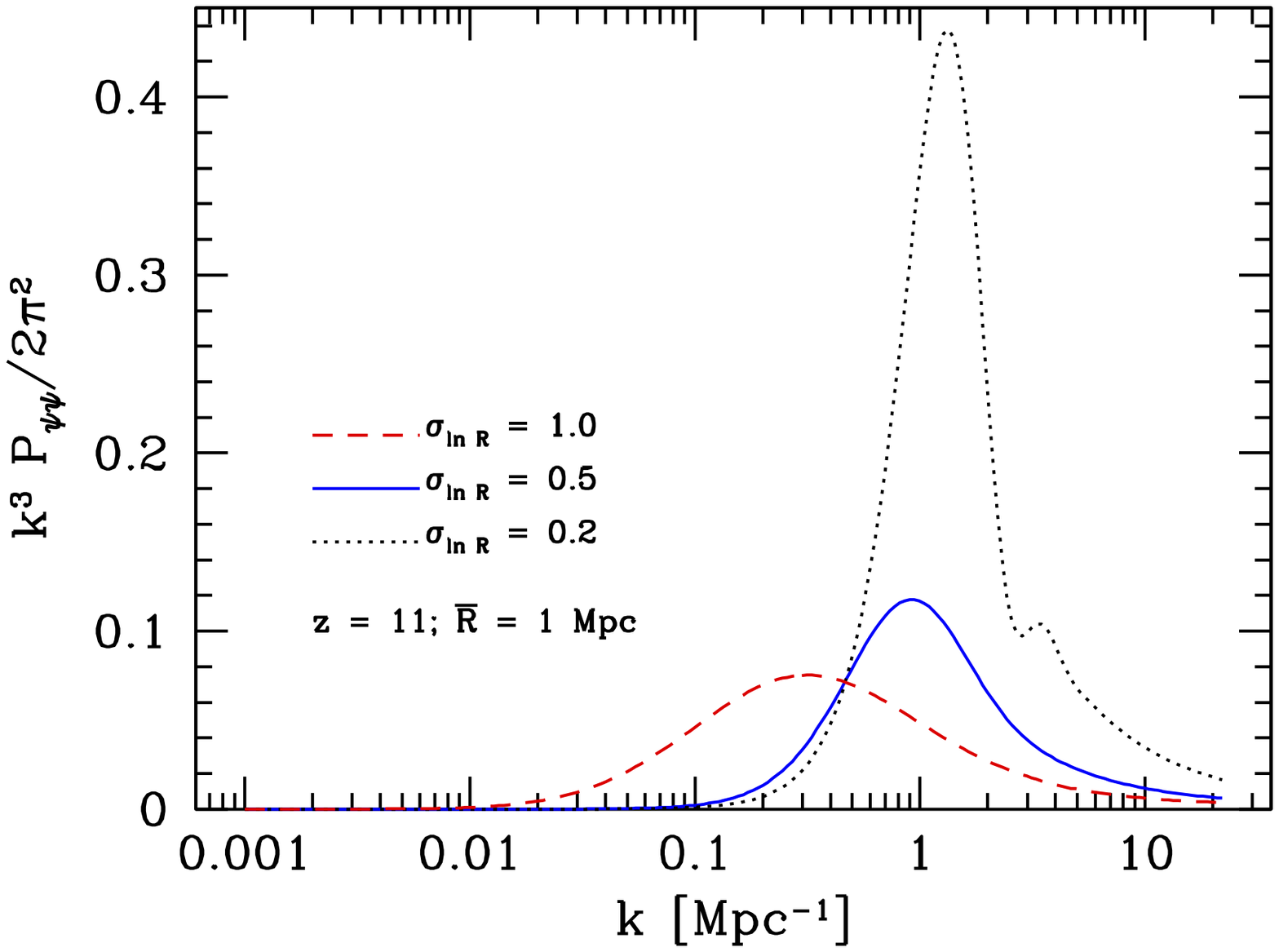}
\caption{$k^3P_{\psi\psi}(k)/2\pi^2$ for $\bar{R}=1$ Mpc, $b=6$ at different values of  $\sigma_{\ln R}$.}
\label{fig:ptotalppvar}
\end{figure}

There is one  caveat, which is that $\tilde{P}_{\delta\delta}$ does not drop as fast as the shot noise, hence at small scales this term can contribute more. 
In fact, it is this term the responsible for the turnover at small scales of the $\tau$-$\tau$ and $21$-$21$ auto power spectra (see Figs. \ref{fig:ptotalXXvar} and \ref{fig:ptotalppvar}).

We remind the reader that the two-bubble contribution to the $X\psi$ cross-correlation is given by:
\be
P^{2b}_{X\psi} &\approx& - \bxh^2 [\ln \bxh b \langle W_R \rangle +1]^2 P_{\delta\delta}(k) +\nonumber \\
&& \bxh [\ln \bxh b \langle W_R \rangle +1] P_{\delta\delta}(k) \nonumber\\
\ee

The minimum value $P^{2b}_{X\psi}$ is reached when $\bxh = e^{-1-1/b}$, while for the shot noise term $\bxe = 0.5$ represents the peak value. We can now compare the two-bubble terms to the shot noise: 
\be
R^{2b-sn}\equiv P^{2b}_{X\psi}/P^{sn}_{X\psi}& \lesssim & \frac{16}{3\pi} b e^{-1-1/b}(1+e^{-1-1/b})\times P_{\delta\delta}(k_{peak}) e^{-27 \sigma_{\ln R} /2}
\ee
For the toy reionization model, we find that $\bar{R}($z=11$)=1.14$ Mpc with a $\sigma_{\ln R}=0.5$ and $b($z=11$) = 4.8$, we find $R^{2b-sn}\sim 0.01$, close to the ratio found in Fig.~\ref{fig:tau21z}. 

Can the two-bubble term ever dominate over the one-bubble term? Since the ratio goes as $1/\bar{R}^3$ and decreases exponentially in $\sigma_{\ln R}$, for bubble distributions with small bubble radius and narrow width, we find that the two-bubble term can easily dominate the total correlation function. When the radius and the variance depend on redshift, we expect the two-bubble term to be increasingly  important to the correlation function in the early stages of reionization, in other words, when the correlation function is dominated by points that live in two different bubbles. At the onset of reionization, the bubbles are small and it is more probable to find two points that live in two separate bubbles. As bubbles merge and grow, it becomes more likely that the correlation function has contribution from points that are in the same bubble. \cite{McQuinn:2005hk} make this distinction, and divide the reionization model in two regimes separated by the average ionization fraction. 

Concluding, we see that the location of the peak of the correlation function is roughly set by $k_{peak}$, given in Eq. \eqref{eq:kpeak}, and we note that the location of the peak is almost equivalent for the one and two-bubble terms. 

We have shown that in the case of a log-normal distribution the contribution from $\tilde{P}_{\delta\delta}$ to the peak amplitude generally is small compared to the shot noise and as such, to the overall correlation. 
Since the shot noise grows as $\propto \bar{R}^3$ and exponentially in the width of the distribution $\sigma_{\ln R}$, we find that assuming smaller values for these parameters lead to rapidly increasing contribution of the two-bubble term compared to the 1-bubble term. Since decreasing both of these parameters also increases the value of $k_{peak}$, assuming a narrower distribution of bubbles with a smaller average radius results in a correlation function that  is dominated by the two-bubble term and peaks at smaller (physical) scales. These  findings are consistent with the expectation that larger bubble imply a larger shot noise. 

 \subsection{Normal distribution}

Simulations show that bubbles are well traced by a log-normal distribution at early times  \citep{Zahn:2006sg}, while at later times the distribution can transition to a normal distribution. Since at late times, large radii dominate reionization, we expect the shot noise to become more dominant.  

A normal distribution is given by
\be
P(R) = \frac{1}{\sqrt{2\pi\sigma^2_{R}}} e^{ [-(R-\bar{R})^2/(2\sigma_{R}^2)]}
\ee
As expected, for a Gaussian, all relevant quantities are much closer to the distribution values (e.g. $\bar{R}$) deviating, by definition, at most $1$ sigma.  

The average bubble volume is given by
\be
\langle V_b \rangle &=& \frac{4}{3} \pi  \bar{R} \left(\bar{R}^2+3 \sigma_R^2\right)
\ee
We will assume that a normal distribution is only valid for $\bar{R}>1$Mpc. In this limit, the above equality holds, even for a bound probability function, as long as $\sigma_R \leq 1$. This cutoff is consistent with observations. The expression above can easily be understood by Wick expanding the 3-point function $\langle R^3 \rangle$.

Similarly, for the variance we obtain:
\be
\langle V_b^2 \rangle &=&\frac{16}{9} \pi ^2 \left(15 \bar{R}^4 \sigma_R^2+45 \bar{R}^2
   \sigma_R^4+\bar{R}^6+15 \sigma_R^6\right)\nonumber\\
\ee

A gaussian distribution allows us to analytically compute the volume average window function: 
\be
\langle W_R\rangle(k)&=&\frac{3 e^{-\frac{1}{2} k^2 \sigma_R^2} \left[\left(k^2 \sigma_R^2+1\right) \sin \left(k\bar{R}\right)-k \bar{R} \cos \left(k
   \bar{R}\right)\right]}{k^3 \left(3 \bar{R} \sigma_R^2+\bar{R}^3\right)},\nonumber \\
\ee

At large scales, 
\be
\lim_{k\bar{R}\ll 1} \langle W_R\rangle = 1,
\ee
while at small scales,
 \be
 \lim_{k \bar{R} \gg 1} \langle W_R\rangle =  \frac{3 \sigma_R^2 e^{-\frac{1}{2} k^2 \sigma_R^2} \sin \left(k
   \bar{R}\right)}{k \bar{R} \left(3 \sigma_R^2+1\right)}
   \ee
   
The variance is given by: 
\
\be
\langle W_R^2\rangle(k)&=&\frac{9 e^{-2 k^2 \sigma_R^2} \left\{e^{2 k^2 \sigma_R^2}
   \left[k^2 \left(\bar{R}^2+\sigma_R^2\right)+1\right]-2 k
  \bar{R} \left(2 k^2 \sigma_R^2+1\right) \sin \left(2 k
   \bar{R}\right)+\left[k^2 \left(\bar{R}^2-3 \sigma_R^2\right)-4
   k^4 \sigma _R^4-1\right] \cos \left(2 k \bar{R}\right)\right\}}{2
   k^6 \left(3 \bar{R} \sigma_R^2+\bar{R}^3\right){}^2} \nonumber\\ 
\ee

At large scales, the variance becomes:
\be
\lim_{k \bar{R}\ll1} \langle W_R^2\rangle = \langle V_b^2\rangle/\langle V_b \rangle^2,
\ee
and at small scales:
\be
\lim_{k \bar{R}\gg1} \langle W_R^2\rangle =9(\bar{R}^2+\sigma_R^2)/(2k^4(\bar{R}^2+3\sigma_R^2)^2
\ee

By equating the two limiting cases, we can derive the peak scale for the variance of $W_R$:

\be
k^{var}_{peak}&=&\left[\frac{9 \left(\bar{R}^2+\sigma_R^2\right) \left(3 \bar{R} \sigma_R^2+\bar{R}^3\right)}{2 \left(15 \bar{R}^4 \sigma_R^2+45 \bar{R}^2 \sigma_R^4+\bar{R}^6+15 \sigma_R^6\right)}\right]^{1/4}\nonumber \\
\ee 
The maximum peak value of the shot noise is then given by
\be
k^3P^{sn}_{X\psi} &\lesssim& (k^{var}_{peak})^3\frac{ \langle V_b^2\rangle}{\langle V_b \rangle}
\ee

A similar approach for the peak scale of the window average does not give an accurate enough answer. Therefore, we will use the following best fit:
\be
k_{peak}^{av} \sim 2.2 \bar{R}^{-1}
\ee
This and the derived maximum value of the shot noise immediately allow us to put a constraint on the ratio between the two-bubble term and the shot noise,
\be
R^{2b-sn}\equiv P^{2b}_{X\psi}/P^{sn}_{X\psi}& \simeq & b\frac{1+\bxh}{1-\bxh} \frac{P_{\delta\delta}(k_{peak}^{av} ) }{\langle V_b\rangle \langle W_R^2 \rangle(k^{var}_{peak})}\nonumber\\
\ee

From this expression, we find that the shot noise is significantly larger than  the two-bubble term around the peak scale. This nicely fits into the previous picture, since the normal distribution of the bubbles is only physical at late time in the reionization history when $\bar{R}>1$ Mpc. At those times the one-bubble term (dominated by the shot noise) should make up the largest contribution to the total cross-correlation spectrum. \\

\bibliography{draft}

\end{document}